\documentclass[12pt,preprint]{aastex}

\begin{document}

\title{\bf PERIODIC ACCELERATION OF ELECTRONS IN THE 1998 NOVEMBER 10 
SOLAR FLARE}
\author{A. Asai \altaffilmark{1,2}, 
M. Shimojo \altaffilmark{3}, H. Isobe \altaffilmark{1,2}, 
T. Morimoto \altaffilmark{1}, T. Yokoyama \altaffilmark{3}, 
K. Shibasaki \altaffilmark{3}, and H. Nakajima \altaffilmark{3}}
\email{asai@kwasan.kyoto-u.ac.jp}
\altaffiltext{1}{
Kwasan and Hida Observatories, Kyoto University, Yamashina-ku, 
Kyoto 607-8471, JAPAN}
\altaffiltext{2}{
Department of Astronomy, Kyoto University, Sakyo-ku, Kyoto 606-8502, JAPAN}
\altaffiltext{3}{
Nobeyama Radio Observatory, Minamisaku, Nagano, 384-1305, JAPAN}

\begin{abstract}

We present an examination of the multi-wavelength observation 
of a C7.9 flare which occurred on 1998 November 10. 
This is the first time of imaging observation of the quasi-periodic 
pulsations (QPPs).
Four bursts were observed with the hard X-ray telescope aboard 
{\it Yohkoh} and the Nobeyama Radioheliograph during the impulsive phase 
of the flare.
In the second burst, the hard X-ray and microwave time profiles clearly 
showed a QPP.
We estimated the Alfv\'{e}n transit time along the flare loop using 
the images of the soft X-ray telescope aboard {\it Yohkoh} and the 
photospheric magnetgrams, and found that the transit time was 
almost equal to the period of the QPP. 
We therefore suggest, based on a shock acceleration model, 
that variations of macroscopic magnetic structures, such as oscillations 
of coronal loops, 
affect the efficiency of particle injection/acceleration.

\end{abstract}


\section{INTRODUCTION}

Non-thermal electrons generated in the impulsive phase of a flare 
are observed in hard X-ray, $\gamma$-ray, and microwaves.
The light curves in these wavelengths show short-lived bursts 
with durations between 10 s and 10 min \citep{Dulk1985}.
These bursts include smaller pulses with shorter duration, 
and they sometimes show periodicity.

A good example of such quasi-periodic pulsations (QPPs) was seen in 
a flare on 1980 June 7 \citep{Kipl1983}.
\citet{Naka1983} and \citet{Kane1983} examined the temporal evolution of 
the X-ray and radio spectra and the spatial structure of the flare.
They suggest that the quasi-periodic pulses indicate a modulation of 
the particle injection/acceleration rate.
\citet{Taji1982} showed by numerical simulation that stored magnetic 
energy is explosively transformed to particle acceleration.
They also suggest that the current loop coalescence instability induces 
the QPPs.
Moreover, \citet{Taji1987} showed that the period of the QPP is equal to 
the Alfv\'{e}n transit time `across' the current loop.
However, these works were mainly based on the total flux, 
and the detailed spatial configuration of the flare that shows QPP is 
still unknown.

On 1998 November 10 a solar flare (C7.9 on the GOES scale) occurred in 
NOAA 8375. 
The flare was observed by {\it Yohkoh} \citep{Oga1991} and 
the Nobeyama radio observatory, and clearly showed quasi-periodic 
behavior in the hard X-ray and microwave time profiles.
In this paper we analyze the QPP using the high resolution X-ray and 
microwave images.
Then we compare the period of the QPP with typical time-scales of flare 
loops.
Finally, we discuss the effect of the magnetic structure on the particle 
injection/acceleration rate.

\section{OBSERVATIONS}

The solar flare occurred in NOAA 8375 (N19, W78) at 00:10~UT, 1998 
November 10.
Microwave images of the flare were taken with the Nobeyama 
Radioheliograph (NoRH; Nakajima et al. 1994) which observed the sun 
at 17 and 34~GHz with a temporal resolution of 1.0~s.
The spatial resolutions of NoRH data are $12^{\prime\prime}$ for 17~GHz 
and $6^{\prime\prime}$ for 34~GHz.
The Nobeyama Radio Polarimeter (NoRP; Torii et al. 1979; Shibasaki 
et al. 1979; Nakajima et al. 1985) measured the total flux of the flare 
at 1, 2, 3.75, 9.4, 17, 34 and 80~GHz with a temporal resolution of 0.1~s.
The hard X-ray images were obtained with the hard X-ray telescope 
(HXT; Kosugi et al. 1991) aboard {\it Yohkoh}, with a spatial and 
temporal resolution of about 5$''$ and 0.5 s, respectively.
The soft X-ray images were also obtained with the {\it Yohkoh} soft X-ray 
telescope (SXT; Tsuneta et al. 1991).
We used full resolution images in the Partial Frame Image mode with 
a spatial resolution of about $2^{\prime\prime}_{\,\cdot} 5$.

Figure 1 shows the microwave, soft X-ray, and hard X-ray time profiles.
The top solid line is that of NoRH 17~GHz, the second dotted line is the 
GOES 1.0 - 8.0~{\AA} channel,
the third solid line is the HXT M1 band (23 - 33~keV), 
and the bottom solid line is the HXT L band (14 - 23~keV).
Four bursts with duration between 10 and 30~s are seen in the microwave 
and hard X-ray emission.
The fine spikes in the second burst clearly show the QPPs in both 
microwaves and hard X-rays. 
In this paper we investigate the quasi-periodic nature of the second 
burst.

Figure 2 shows the images of NoRH/17~GHz, the HXT/M1~band, and SXT 
at the second burst.
The flare shows a double source in the microwave data 
({\it gray contours}).
We refer to the northern and brighter source as the source ``A'' and 
the other one as source ``B''.
On the other hand, in the hard X-ray data ({\it black contours}) 
only one source is seen near source ``B''. 
This is common in the L and M1 bands of HXT, but it is unknown whether 
it is also common in the M2 (33 - 53~keV) and H (53 - 93~keV) bands.
We could not synthesize the hard X-ray images in these bands at the time 
due to insufficient counts.
The soft X-ray image ({\it gray scale}) shows a small bright kernel at 
the same region at source ``B''. 
We can also see a faint loop which connects the bright kernel and the 
radio source ``A''. 

\section{PERIODIC PULSATION}

Figure 3 presents the hard X-ray and microwave time profiles for the 
second burst, from 00:14:20~UT to 00:14:40~UT.
The QPPs are clearly seen.
In the time profiles obtained with the NoRP data the QPPs are seen in 
9.4 and 17~GHz.
The period calculated from the autocorrelation function of the time 
profiles was 6.6~s.
We also calculated the period of the QPPs in hard X-rays 
(M1 and M2 bands), and found that the period (6~s) was almost the same 
as those in NoRP.
We analyzed the microwave spectrum using the NoRP data and found that 
the power-law distribution index was about $-2 \sim -1.5$ and the 
turn-over frequency was about 10~GHz.
These results indicate that the emission of the QPP originated from 
gyrosynchrotron radiation from non-thermal electrons.
Since both the time profiles of the M1 and M2 bands clearly showed the 
QPPs, the emission in hard X-rays was also produced by non-thermal 
electrons.

As described in section 2, the flare has two microwave sources (see Figure 2). 
We found that the time profile of source ``A'' showed the QPP, 
while that of source ``B'' did not.
The pulsating source is different from the hard X-ray source 
that is located near source ``B''.
To clarify the relation between the microwave pulsation at source ``A'' 
and the hard X-ray pulsation, 
we calculated the correlation between the time profiles of 17~GHz 
and the M1 band as a function of the time lag between them.
The maximum correlation was found for a delay of the microwave pulsation 
with a time of 0.6 -- 1.0~s. 

We suggest that the acceleration site lies near the hard X-ray source, 
and that the delay time between microwaves and hard X-rays is explained 
by the flight duration of non-thermal electrons between both the sources.
The length of the soft X-ray faint loop connecting the hard X-ray source 
and source ``A'' is about $5.0 \times 10^4$ km.
If we assume that the velocity of the electrons is about 
$1.0 \times 10^{5}$~{km s$^{-1}$} ($\sim$ 30 \% of the speed of light), 
the delay time between microwaves and hard X-rays is about 0.5~s, 
which is almost equal to the observed delay time.
The soft X-ray images show two parts related to the flare: 
a bright kernel near source ``B'' and a faint large loop 
connecting the kernel and source ``A'' (see Figure 2).
We believe that the flare occurred at the contact region of the two parts 
via magnetic reconnection \citep{Hana1997,Hey1977}.
In such a scenario, the acceleration site would be located near the 
reconnection site.
Hence, it is reasonable to suggest that the acceleration site is located 
near source ``B'' and the hard X-ray source.

Here, another question occurs; 
why did the time profile of the microwave source ``B'' not show the QPP? 
As an answer, we suggest that the emission mechanism of source``B'' differs 
from that of source ``A''.
The emission mechanism of source ``A'' is optically-thin gyrosynchrotron 
radiation from non-thermal electrons.
On the other hand, the dominant emission mechanism of source ``B'' is 
optically-thin thermal Bremsstrahlung since the spectrum index of source 
``B'' that is derived from the 17 and 34~GHz is about 0.
Therefore, the time profiles of source ``B'' do not show the population 
of the accelerated/injected non-thermal electrons.
To confirm the suggestion, we estimated the thermal radio flux from source 
``B'' by using the temperature and the emission measure of SXT bright 
kernel (see section 4), and found that the flux was almost the same 
as the observed flux of source ``B''.
We also suggest that the thermal emission was radiated from the thermal 
plasma that was confined by the flare loops.
The loops were formed at the first burst via magnetic reconnection.
Therefore, although the fine structure of the kernel is not resolved in 
the SXT images, we think that it consists of the main flare loops 
(see Figure 2).

\section{TYPICAL TIMESCALES OF FLARE LOOP}

In the previous section we showed that the period of the pulsation was 6.6~s.
What is the explanatin for the period?
To answer the question, we derive some typical time-scales of the flare loop 
and discuss them.
Temperature, density, size, and magnetic field strength of the flare loop 
are needed for the estimation.
The temperature $T$ and the volume emission measure $EM$ are derived 
from the soft X-ray images which were taken with the SXT thick aluminum (Al12)
and beryllium (Be) filters.
The region of flare loop $A$ is defined as the region where the intensity 
of an SXT Be filter image is more than 2,000~DN s$^{-1}$ pix$^{-1}$.
This leads to an area of 2.5$\times10^{18}$~cm$^2$.
Using the filter ratio method \citep{Hara1992}, 
the average temperature and volume emission measure over the flare loop are 
calculated to be 9.4~MK and 8.1$\times$10$^{48}$~cm$^{-3}$, respectively.

Here, we assume that the kernel is a cube, and that the volume $V$ of 
the flare kernel is the area to the power of three halves, 
i.e. $V = A^{3/2}$.
Then, the number density $n$ in the top of the flare loop becomes 
$n = \sqrt{EM/V} = \sqrt{EM/A^{3/2}} \approx 4.5\times10^{10} 
\;\mathrm{cm^{-3}}$.
Also, the acoustic velocity $c_s$ in the flare loop is given by 
$c_s \approx \sqrt{\gamma k_B T / m_p} \approx 360 \;\mathrm{km \;s^{-1}}$, 
where $k_B$ is Boltzmann's constant; $m_p$ is the proton mass; 
$\gamma$ is the ratio of specific heats and is here assumed to be $5/3$.
Since the structure of the flare loop is not resolved in the SXT images, 
we can not measure the length $l$ and width $w$ of the flare 
loop correctly. 
Therefore, we use the square root of the area instead of the length of 
the flare loop ($l \sim A^{1/2} \sim $ 16,000~km).
Then, the acoustic transit time along the flare loop is 
$\tau_{sl} = l/c_s \approx 44 \;\mathrm{s}$.
On the other hand, if we use the width of the faint loop instead of 
the width of the flare loop ($w \sim $ 6,000~km), 
the acoustic transit time across the flare loop becomes 
$\tau_{sw} = l/c_s \approx 17 \;\mathrm{s}$.
Both the time-scales are much longer than the observational period of 
the QPP (6.6~s).

In order to derive the Alfv\'{e}n transit times of the flare loop, 
we have to estimate the magnetic field strength in the corona.
However, this estimation is difficult in our case as; 
(1) we can not measure the magnetic field strength of the corona directly, 
(2) we can not obtain the actual magnetic field strength 
in the photosphere at the flare region since it is located near 
the north-west limb.
Therefore, we calculate the potential field of the flare region based on 
a magnetogram on November 6.
The magnetogram was obtained with the Michelson Doppler Interferometer 
(MDI; Scherrer et al. 1995) aboard the Solar and Heliospheric Observatory 
({\it SOHO}; Domingo et al. 1995).
The field strength on the apex of the flare loop $B$ is estimated to be 
about 300~G using the potential field extrapolation by the software package 
MAGPACK2 \citep{Saku1982}.
As a result, the Alfv\'{e}n velocity $c_A$ in the flare loop becomes 
$c_A \approx B / \sqrt{4\pi n m_p} \approx 3,100 \;\mathrm{km \;s^{-1}}$.
Then, the Alfv\'{e}n transit time `along' the flare loop is 
$\tau_{Al} = l/c_A \approx 5.1 \;\mathrm{s}$ 
and the Alfv\'{e}n transit time `across' the flare loop is 
$\tau_{Aw} = w/c_A \approx 1.9 \;\mathrm{s}$.

The values of our estimations are summarized in Table 1. 
Although the Alfv\'{e}n transit time is sensitive to the magnetic field 
strength estimated above, 
the transit time along the flare loop ($\tau_{Al}$) is the most similar to 
the observational period of the pulsation in microwave and hard X-rays.

\section{DISCUSSION}

We investigated the impulsive phase of the flare whose time evolution 
showed a clear QPP in microwaves and hard X-rays. 
We found that the periods of the QPP were 6.6 s and 
that the acceleration site of non-thermal electrons was located near 
source ``B'' using correlation analysis of the time profiles.
Then, we estimated some typical time-scales of the flare loop from the 
observational data and found that the Alfv\'{e}n transit time along the 
flare loop was close to the period of the pulsation. 

\citet{Taji1987} suggest that the periods of QPP in microwaves and hard 
X-rays are equal to an Alfv\'{e}n transit time `across' the flare loop 
($\tau_{Aw}$).
However, this transit time was much shorter than the observational period of 
the QPP.
This implies that the origin of the periodic pulsation in our case is not 
the coalescence instability in the reconnection site.

Recently, {\it Transition Region and Coronal Explorer} ({\it TRACE}; 
Handy et al. 1999) observations have shown coronal loop oscillations 
that was induced by a flare \citep{Naka1999}.
T. Miyagoshi (in preparation) performed three-dimensional MHD simulations 
of the coronal loop oscillation and found that the period of the loop 
oscillation is equal to the Alfv\'{e}n transit time `along' the oscillating 
loops ($\tau_{Al}$).
Does the oscillation of the coronal loop relate with the particle 
acceleration/injection?
\citet{Tsu1998} propose that non-thermal electrons can efficiently be 
accelerated by a first order Fermi process at the fast shock 
located below the reconnection X-point.
They suggest that the accelerated electrons are trapped between the two 
slow shocks, and that the energy injection depends on the length of 
the fast shock which is pinched by these slow shocks.
If the reconnected (flare) loop that is located under the fast shock is 
oscillating, then the length of the fast shock probably varies with the 
loop oscillation, synchronously.
Hence, we propose, under the hypothesis of the acceleration model proposed 
by Tsuneta and Naito (1998), 
that the origin of the QPP in microwaves and hard X-rays 
is the modulation of the acceleration/injection of non-thermal electrons.
Moreover, we propose that the modulation is produced by the variations of 
macroscopic magnetic structures, for example, oscillations of coronal loops.

\acknowledgments
We first acknowledge an anonymous referee for his/her useful comments and 
suggestions, which improved this paper.
We'd like to thank Drs. K. Shibata, T. Terasawa, T. Saito, T. T. Ishii, and 
T. Miyagoshi for fruitful discussions.
We also thank Drs. D. H. Brooks, T. Miyagoshi for their careful reading 
and helpful comments.
We are indebted to Dr. T. Sakurai for allowing us to use his potential 
field program (MAGPACK2).
We also thank all the members of Nobeyama Radio observatory for 
their support in our observation and study.
We made extensive use of {\it SOHO} MDI Data Service and {\it Yohkoh} 
Data Center.


\begin{table}
\begin{center}
Table 1: Physical values of flare loop  \\

\begin{tabular}{lrr}\tableline\tableline
area ($A$)                  & 2.5$\times 10^{18}$ & [cm$^2$] \\
$\;\;$volume ($V$)          & 4.0$\times 10^{27}$ & [cm$^3$] \\
$\;\;$length ($l$)          & 1.6$\times 10^{9}$  & [cm] \\
$\;\;$width ($w$)    & $\sim$ 6.0$\times 10^{8}$  & [cm] \\
temperature ($T$)           & 9.4$\times 10^{6}$  & [K] \\
number density ($n$)        & 4.5$\times 10^{10}$ & [cm$^{-3}$ ] \\
magnetic field ($B$)        & $\sim$ 300          & [G] \\ \tableline
acoustic velocity ($c_s$)   & 3.6$\times 10^{2}$  & [km s$^{-1}$] \\
Alfv\'{e}n velocity ($c_A$) & 3.1$\times 10^{3}$  & [km s$^{-1}$] \\  
\tableline
acoustic transit time \\
$\;\;$along the loop ($\tau_{sl}$)    & 44  & [s] \\
$\;\;$across the loop ($\tau_{sw}$)   & 17  & [s] \\
Alfv\'{e}n transit time \\
$\;\;$along the loop ($\tau_{Al}$)  & 5.1 & [s] \\
$\;\;$across the loop ($\tau_{Aw}$) & 1.9 & [s] \\ \tableline

\end{tabular}
\end{center}
\end{table}

\clearpage

\begin{figure}
\plotone{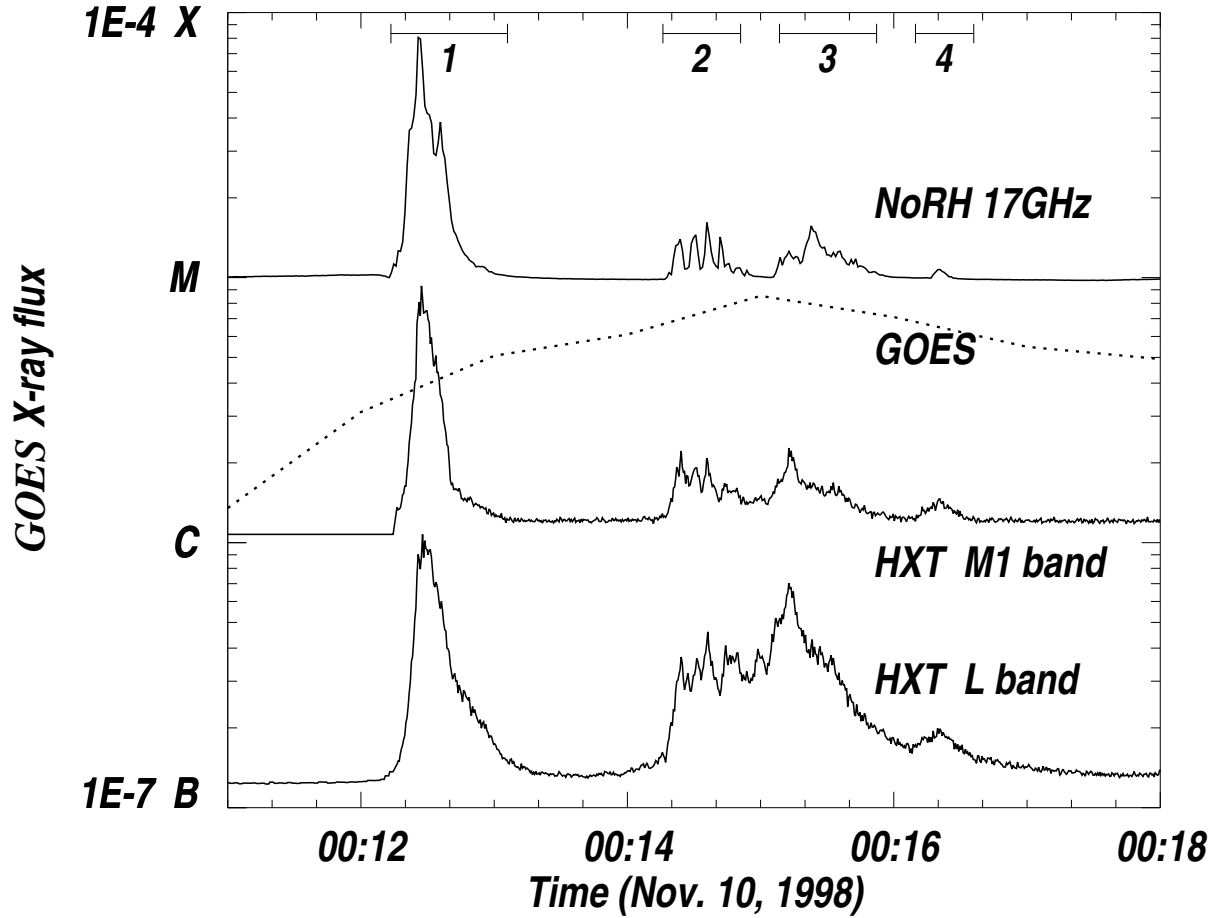}
\caption{
Temporal evolution of the 1998 November 10 flare.
From top to bottom: radio correlation plot observed at 17 GHz with NoRH 
(scaled arbitrarily); 
soft X-ray flux in the GOES 1.0 - 8.0 {\AA} channel ({\it dotted line}); 
hard X-ray count rate measured in the M1 band (23 - 33 keV) and 
L band (14 - 23 keV) of {\it Yohkoh}/HXT (scaled arbitrarily).
Four bursts are identified by the numbered top bars.
\label{fig1}}
\end{figure}

\begin{figure}
\plotone{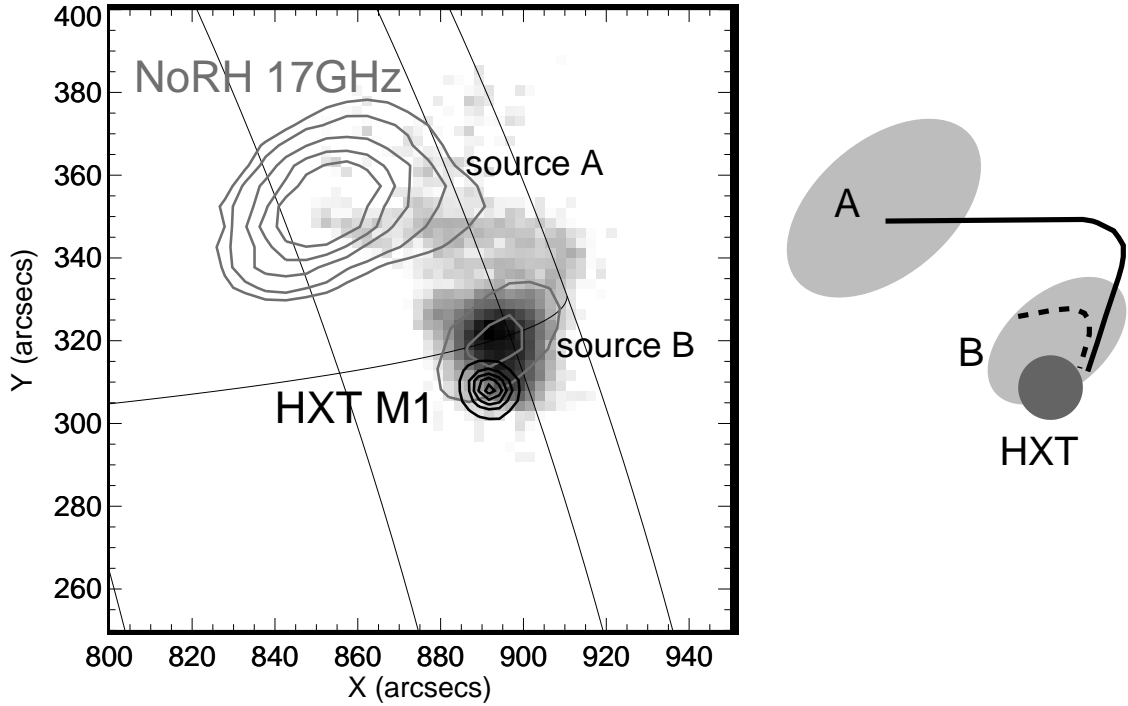}
\caption{Left: coaligned images of the flare.
SXT (Be filter) image is shown by gray scale at 00:14:32 UT.
NoRH (17 GHz; {\it gray contour}) 
image of 00:14:37 UT is plotted at brightness temperatures of 
150,000, 200,000, 300,000, 400,000, and 500,000 K.
HXT (M1 band, 23 - 33 keV) image at 00:14:33 UT is drawn in black contour.
Contour levels are 90, 70, 50, 30, and 10 \% of the peak intensity.
Right: a simple sketch of this region.
The NoRH sources, the HXT source, and SXT faint loop are displayed in 
light gray, dark gray, and black solid line, respectively.
There probably are flare loops (black broken line; see section 3) in 
microwave source ``B''.
\label{fig2}}
\end{figure}

\begin{figure}
\plotone{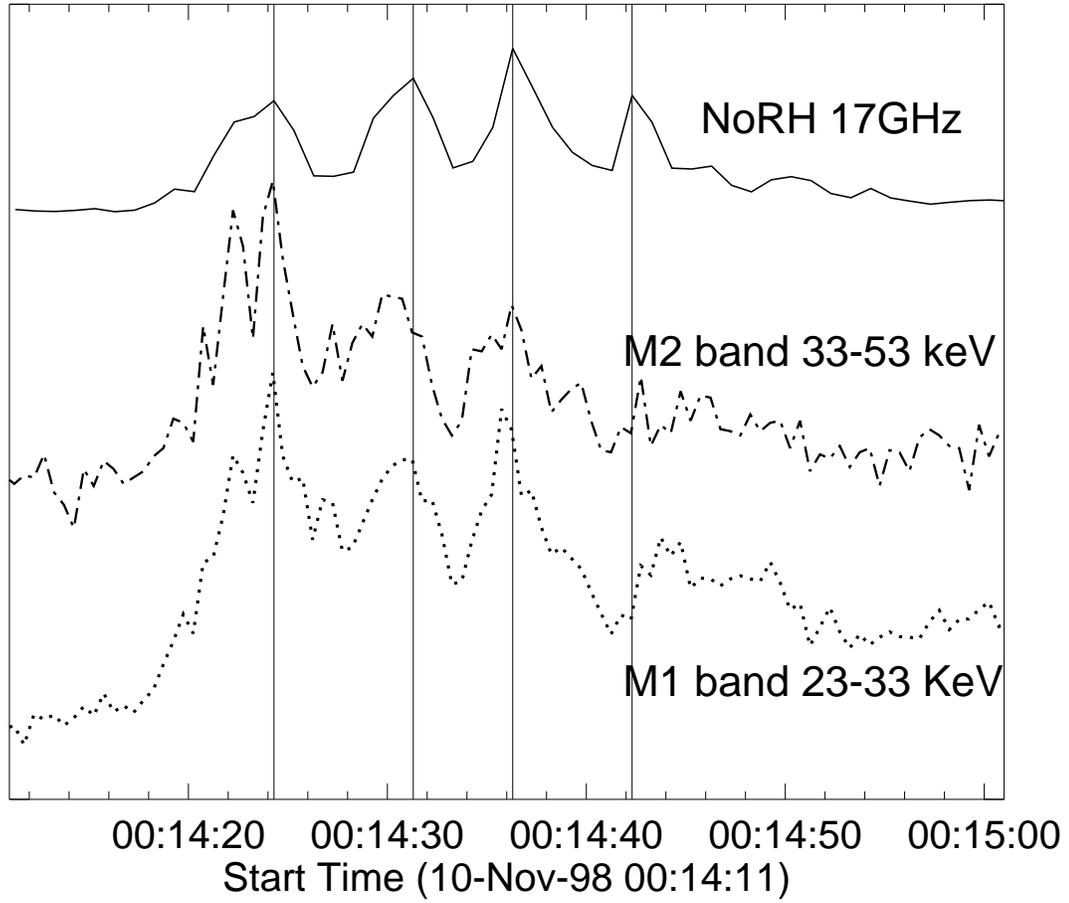}
\caption{
Light curves of the second burst (scaled arbitarily).
From top to bottom: radio brightness temperature observed at 17 GHz by NoRH 
({\it solid line}); 
hard X-ray count rate measured in the M2 band (33 - 53 keV) 
({\it dash dot line}) and 
M1 band (23 - 33 keV) ({\it dotted line}) of {\it Yohkoh}/HXT.
The vertical lines show the peak times of the microwave emission.
\label{fig3}}
\end{figure}

\end{document}